\documentstyle[seceq,mbf,wrapft]{ptptex}
\def\lsim{\mathrel{\rlap{\lower4pt\hbox{\hskip1pt$\sim$}}
    \raise1pt\hbox{$<$}}}                % less than or approx. symbol
\def\gsim{\mathrel{\rlap{\lower4pt\hbox{\hskip1pt$\sim$}}
    \raise1pt\hbox{$>$}}}                % greater than or approx. symbol

\newcommand{\epa}{\varepsilon_{\mu\alpha\nu\beta}}

\newcommand{\Sa}{S_{\mu\alpha\nu\beta}}

\pubinfo{}  %Editorial Office use
\preprintnumber[3cm]{%<-- [..]: optional width of preprint # column.
KUNS-1325\\PTPTeX ver.0.8\\ August, 1997}
\title{%        %You can use \\ for explicit line-break
On the Relations Between Polarized  Structure Functions}
\author{%       %Use \sc for the family name
Johannes {\sc Bl\"umlein}}
\inst{%         %Affiliation, neglected when [addenda] or [errata]
Deutsches Elektronen-Synchrotron, \\
DESY Zeuthen, Platanenallee 6,
D--15735 Zeuthen, Germany
}

\recdate{%      %Editorial Office will fill in this.
December 30, 1999}

\abst{%         %this abstract is neglected when [addenda] or [errata]
\noindent
The status of the twist--2 and the twist--3 integral relations between
polarized structure functions in deep inelastic scattering for
electromagnetic and electroweak interactions is reviewed. One novel
integral relation for the twist--3 contributions can be tested in the 
upcoming high statistics measurements of the structure function
$g_2(x,Q^2)$ in the range $Q^2 \gsim M_p^2$.}

\begin{document}

\maketitle

%\tableofcontents

\makeatletter
\if 0\@prtstyle
\def\asp{.3em} \def\bsp{.26em}
\else
\def\asp{.3em} \def\bsp{.3em}
\fi \makeatother

%%%%%%%%%%%%%%%%%%%%%%%%%%%%%%%%%%%%%%%%%%%%%%%%%%%%%%%%%%%%%%%%%%%%%%%%
\section{Introduction}
%%%%%%%%%%%%%%%%%%%%%%%%%%%%%%%%%%%%%%%%%%%%%%%%%%%%%%%%%%%%%%%%%%%%%%%%

\vspace{1mm}
\noindent
The measurement of the nucleon structure functions in polarized deeply 
inelastic scattering reveals the behaviour of quarks and gluons in
outer magnetic fields and allows to test basic predictions of Quantum
Chromodynamics (QCD)
in the short--distance regime $Q^2 \gg M_p^2$. In this
domain
the structure of the hadronic tensor $W_{\mu\nu}$ which describes the 
process
can be investigated in terms of the light--cone expansion~\cite{LC}.
We restrict our consideration to those contributions to $W_{\mu\nu}$
which contribute to the scattering cross sections in the massless
quark case. There $W_{\mu\nu}$ is a current conserving quantity for
the electro--weak interactions,
%----------------------------------------------------------------------
\begin{eqnarray}
W_{\mu\nu} =&& \left(-g_{\mu\nu}+\frac{q_\mu q_\nu}{q^2}\right)F_1(x,Q^2)
  + \frac{\hat P_\mu\hat P_\nu}{P.q} F_2(x,Q^2)
-i\varepsilon_{\mu\nu\lambda\sigma}\frac{q^\lambda P^\sigma}{2P.q}
F_3(x,Q^2)
\nonumber \\
&& +i\varepsilon_{\mu\nu\lambda\sigma}\frac{q^\lambda S^\sigma}{P.q}
g_1(x,Q^2) 
+i\varepsilon_{\mu\nu\lambda\sigma}
\frac{q^\lambda (P.q S^\sigma-S.q P^\sigma)}{(P.q)^2}
g_2(x,Q^2) \nonumber \\
&&+\left[\frac{\hat P_\mu\hat S_\nu+\hat S_\mu\hat P_\mu}{2} -
S.q\frac{\hat P_\mu\hat P_\nu}{P.q}\right] \frac{g_3(x,Q^2)}
{P.q}\nonumber \\
&&+S.q\frac{\hat P_\mu\hat P_\nu}{(P.q)^2} g_4(x,Q^2) +
\left(-g_{\mu\nu}+\frac{q_\mu q_\nu}{q^2}\right)\frac{(S.q)}{P.q}
g_5(x,Q^2),
\label{had1}
\end{eqnarray}
%--------------------------------------------------------------------------
with
%--------------------------------------------------------------------------
\begin{eqnarray}
\hat P_\mu = P_\mu-\frac{P.q}{q^2}q_\mu,~~~~~~~~~~~~~~~~~
\hat S_\mu = S_\mu-\frac{S.q}{q^2}q_\mu~.
\nonumber
\end{eqnarray}
%--------------------------------------------------------------------------
Here $P,q$ and $S$ are the 4--vectors of the nucleon momentum,
the
momentum transfer and the
nucleon spin, respectively. The unpolarized
structure functions are denoted by $F_i(x,Q^2)$ and the polarized
structure functions by $g_i(x,Q^2)$. $W_{\mu\nu}$ refers to the 
contributions per current, where
the propagator terms are separated off.
In the case of electromagnetic interactions only the structure functions
$F_1, F_2, g_1$ and $g_2$ contribute. The structure functions aquire
their $Q^2$ dependence as scaling violations due to higher order
QCD corrections and power corrections, such as target mass corrections
and dynamical higher twist terms. The latter contributions turn out to be
particularly essential, even if QCD corrections are not yet considered,
in the range of lower values of $Q^2$, $M_p^2 \lsim Q^2$, where most of 
the present data are taken.
%%%%%%%%%%%%%%%%%%%%%%%%%%%%%%%%%%%%%%%%%%%%%%%%%%%%%%%%%%%%%%%%%%%%%%%%
\section{Twist Decomposition}
%%%%%%%%%%%%%%%%%%%%%%%%%%%%%%%%%%%%%%%%%%%%%%%%%%%%%%%%%%%%%%%%%%%%%%%%

\vspace{1mm}
\noindent
The Fourier transform of the Compton amplitude $T_{\mu\nu}^{NC}(x)$
for neutral current interactions reads
%----------------------------------------------------------------------
\begin{eqnarray}
i\int d^4x e^{iqx}  \widehat
T_{\mu\nu}^{NC} &=&
 -  \int\frac{d^4k}{(2\pi)^4}
\overline U(k)\gamma_\mu(g_{V_1}+g_{A_1}\gamma_5)
\frac{\hat k+\hat q    }{(k+q)^2      }
\gamma_\nu(g_{V_2}+g_{A_2}\gamma_5)       U(k)\nonumber\\
& & -\int\frac{d^4k}{(2\pi)^4}
\overline U(k)\gamma_\mu(g_{V_1}+g_{A_1}\gamma_5)
\frac{\hat k-\hat q    }{(k-q)^2      }
\gamma_\nu(g_{V_2}+g_{A_2}\gamma_5)    U(k), \nonumber\\
&=& -i(g_{V_1}g_{V_2}+g_{A_1}g_{A_2})\epa 
q^\alpha u_+^\beta \nonumber\\ & &
+
(g_{V_1}g_{A_2}
+g_{A_1}g_{V_2})\Sa [q^\alpha u_-^\beta 
+u^{\alpha\beta}],
\label{FourTNC}
\end{eqnarray}
%----------------------------------------------------------------------
where $U(k)=\int{       d^4k/ (2\pi)^4  e^{-ikx}\psi(x)}$,
$S_{\mu\alpha\nu\beta}   =   g_{\mu\alpha}g_{\mu\beta}
+g_{\mu\beta}g_{\nu\alpha}-g_{\mu\nu}g_{\alpha\beta}$, and
%--------------------------------------------------------------------------
\begin{eqnarray}
u^\beta_{\pm} &=& -
\int\frac{d^4k}{(2\pi)^4}\overline U(k)
\frac{\gamma_\beta \gamma_5}{(k+q)^2}U(k)~
\mp ~(q\leftrightarrow-q)~,\\
u^{\alpha\beta} & =& -
\int\frac{d^4k}{(2\pi)^4}\overline U(k)
\frac{k_\alpha\gamma_\beta \gamma_5}{(k+q)^2}U(k)~
- ~(q\leftrightarrow-q)~.
\end{eqnarray}
%-------------------------------------------------------------------------
The expansion of the denominators $(k \pm q)^2$ results into the
operator--representation
%--------------------------------------------------------------------------
\begin{eqnarray}
u^\beta_{\pm} &=&
\sum_{n~even,odd}{\biggl(\frac{2}{Q^2}\biggr)^{n+1} q_{\mu_1}\ldots 
q_{\mu_n}
\Theta^{+\beta\{\mu_1\cdots\mu_n\}}}~,\label{ub}\\
u^{\alpha\beta} &=&
\sum_{n~even}{\biggl(\frac{2}{Q^2}\biggr)^{n+1} q_{\mu_1}\ldots 
q_{\mu_n}
\Theta^{+\beta\{\alpha\mu_1\cdots\mu_n\}}}~.
\label{uab}  
\end{eqnarray}
%--------------------------------------------------------------------------
The operators  $\Theta^{+\beta\{\mu_1\cdots\mu_n\}}$ are given by
%--------------------------------------------------------------------------
\begin{eqnarray}
\Theta^{+\beta\{\mu_1\cdots\mu_n\}} =
\int{\frac{d^4k}{(2\pi)^4} \overline U(k)\gamma_\beta\gamma_5
k_{\mu_1}\ldots k_{\mu_n}       U(k)} =
\Theta^{+\beta\{\mu_1\cdots\mu_n\}}_S +
\Theta^{+\beta\{\mu_1\cdots\mu_n\}}_R  \nonumber\\
\end{eqnarray}
%--------------------------------------------------------------------------
and may be decomposed into a fully
symmetric and a remainder part with respect to their indices. 
The former
contribution is of twist--2 while the latter is a twist--3 operator.
In general this decomposition has to be performed accounting for target
mass effects, see Ref.~\cite{BT} for details. In the massless case the 
corresponding representations were given in Ref.~\cite{BK1}\footnote{
For the twist--2 contributions one may obtain   representations
 using
the covariant parton model~\cite{BK2} in the {\it massless} case and
lowest order
in $\alpha_s$.}.

Whereas the above decomposition of the Compton amplitude yields 
contributions to all the eight Lorentz tensors of Eq.~(1) due to
$\Theta^{+\beta\{\mu_1\cdots\mu_n\}}_S$
this is the case for the polarized part for
$\Theta^{+\beta\{\mu_1\cdots\mu_n\}}_R$ only keeping all the nucleon
mass terms~\cite{BT}. Taking the limit $M_p \rightarrow 0$ only 
contributions $ \propto g_2$ and $g_3$ are obtained, which were studied
in Ref.~\cite{BK1} previously.

The deep inelastic  structure functions contain contributions of different
twist. Since the individual twist terms obey independent renormalization
group equations their scaling violations are different. Moreover the
different twist operators aquire separate expectation values which are
related to different target mass corrections in general. Because of this
the structure functions $F_i$ and $g_i$ have to be represented as linear
superpositions of their twist contributions, the scaling violations of
which are calculated individually. Henceforth we will discuss the
twist contributions separately.
%%%%%%%%%%%%%%%%%%%%%%%%%%%%%%%%%%%%%%%%%%%%%%%%%%%%%%%%%%%%%%%%%%%%%%%%
\section{Twist--2 Relations}
%%%%%%%%%%%%%%%%%%%%%%%%%%%%%%%%%%%%%%%%%%%%%%%%%%%%%%%%%%%%%%%%%%%%%%%%

\vspace{-1mm}
\noindent
%-----------------------------------------------------------------------
In lowest order in $\alpha_s$ the twist--2 contributions to the
structure functions $\left. g_i\right|_{i=1}^5$ are determined by a
single quarkonic operator matrix element $a_n$ for each moment $n$
in the massless case
%-----------------------------------------------------------------------
\begin{equation}
g_i(n) = \int_0^1 dx x^{n-1} g_i(x)~.
\end{equation}
%-----------------------------------------------------------------------
These non--perturbative functions are different for the sets of structure
functions $g_{1,2}$ and $g_{3,4,5}$ due to  the corresponding quark
contents. At the level of twist--2 the former ones are 
$\propto \Delta q(x) + \Delta \overline{q}(x)$
while the latter are of the type
$        \Delta q(x) - \Delta \overline{q}(x)$.

The nucleon mass dependence may induce involved expressions for the 
different structure functions which are typically of the form, 
cf.~\cite{BT},\footnote{Numerical results are presented in 
Ref.~\cite{BT1}.}
%-----------------------------------------------------------------------
\begin{eqnarray}
g_{3~\tau=2}^{\pm}(x) &=&
\sum_q {~g_V^qg_A^q 
\Biggl\{
\frac{2x^2}{\xi(1+4 M^2 x^2/Q^2)^{3/2}}
\int_{\xi}^{1}{\frac{d\xi_1}{\xi_1}F^{\pm q}(\xi_1)}}\nonumber\\  
& & +
\frac{12 M^2 x^3/Q^2}{(1+4 M^2 x^2/Q^2)^2}
\int_{\xi}^{1}{\frac{d\xi_1}{\xi_1}\int_{\xi_1}^{1}{
\frac{d\xi_2}{\xi_2}F^{\pm q}(\xi_2)}}\nonumber \\
& & +
\frac{3}{2}\frac{(4 M^2 x^2/Q^2)^2}{(1+ 4 M^2 x^2/Q^2)^{5/2}}
\int_{\xi}^{1}{d\xi_1\int_{\xi_1}^{1}{
\frac{d\xi_2}{\xi_2}\int_{\xi_2}^{1}{\frac{d\xi_3}{\xi_3}F^{\pm q}
(\xi_3)}}}
\Biggr\}~.
\label{g3x}
\end{eqnarray}
%------------------------------------------------------------------------
Despite of this one may show, however,  that the
relations
%------------------------------------------------------------------------
\begin{eqnarray}
\label{eqg2}
g_2^{\rm II}(x,Q^2) &=&- g_1^{\rm II}(x,Q^2)
+ \int_x^1 \frac{dy}{y}  g_1^{\rm II}(y,Q^2) \\
g_3^{\rm II}(x,Q^2) &=&  2x
  \int_x^1 \frac{dy}{y^2}   g_4^{\rm II}(y,Q^2)
\label{eqg3}
\end{eqnarray}
%------------------------------------------------------------------------
hold {\it in general}~\cite{BT}
\footnote{The validity of Eq.~(\ref{eqg2}) in the presence of
target mass corrections was shown in \cite{PR} before. Expansions in terms
of $M^2/Q^2$ as considered in this paper, however, turn out to
introduce artificial singularities in the range $x \rightarrow 1$, which
are not present in the resummed expressions~\cite{BT}.}.
In this way the target mass corrections are absorbed, without changing
the form of the Wandzura--Wilczek relation  (\ref{eqg2})~\cite{WW}
and a relation by the author and Kochelev~\cite{BK1,BK2}  
(\ref{eqg3}) in the massless case.
As was shown in Ref.~\cite{BT} the Wandzura--Wilczek relation also holds
in the case that the quark mass terms are considered as well.

The relation
%------------------------------------------------------------------------
\begin{eqnarray}
g_4^{\rm II}(x,Q^2)  =   2x g_5 ^{\rm II}(x,Q^2) +\Delta(M^2/Q^2)
\end{eqnarray}
%------------------------------------------------------------------------
with $\lim_{\rho \rightarrow 0} \Delta(\rho) = 0$ turns out to be a
relation between structure functions only in the massless case, where
it was found by Dicus~\cite{DIC}. In the presence of target masses
it is modified~\cite{PG}
 in the same way as the Callan--Gross relation~\cite{CG}.
%%%%%%%%%%%%%%%%%%%%%%%%%%%%%%%%%%%%%%%%%%%%%%%%%%%%%%%%%%%%%%%%%%%%%%%%
\section{Why do integral relations occur?}
%%%%%%%%%%%%%%%%%%%%%%%%%%%%%%%%%%%%%%%%%%%%%%%%%%%%%%%%%%%%%%%%%%%%%%%%

\vspace{-1mm}
\noindent
%-----------------------------------------------------------------------
The Callen--Gross or Dicus relation and the
Wandzura--Wilczek relation are structurally
different. Whereas the first one is a purely algebraic relation the second
one contains
an integral term. The structure function under the integral 
$g_1^{\rm II}(x,Q^2)$ has a partonic interpretation as well as the
twist--2 contributions to the unpolarized structure function
$F_1^{\rm II}(x,Q^2)$. This difference can be understood studying  
firstly deeply--virtual non--forward Compton scattering $\gamma_1^*(q_1)
+ p_1 \rightarrow \gamma_2^*(q_2) + p_2$ and performing then the limit to
the case of forward scattering in the Compton amplitude~\cite{BGR}
%------------------------------------------------------------------------
\begin{eqnarray}
\lim_{p_- \rightarrow 0}
T_{\mu\nu}(q,p_+,p_-) =  T_{\mu\nu}^{\rm forw.}(p,q)~,
\end{eqnarray}
%------------------------------------------------------------------------
with $p_{\pm} = p_2 \pm p_1, p = p_+/2$ and $q = (q_1 + q_2)/2$. The
Compton amplitude has the representation
%------------------------------------------------------------------------
\begin{eqnarray}
T_{\mu\nu}(q,p_+,p_-) =
T_{\mu\nu}^{\rm sym}(q,p_+,p_-) +
T_{\mu\nu}^{\rm asym}(q,p_+,p_-)~.
\end{eqnarray}
%------------------------------------------------------------------------
From the non--forward expressions~\cite{BGR} one obtains in the
forward limit~:
%------------------------------------------------------------------------
\begin{eqnarray}
T_{\mu\nu}^{\rm sym}(q,p) &=& \left(g_{\mu\nu} - \frac{p_{\mu} q_{\nu}
+p_{\nu}q_{\mu}}{p.q}\right) \int Dz
\nonumber \\   & &~~~~
\Biggl\{\left(\frac{1}{\xi + z_+ - i \varepsilon} -
             \frac{1}{\xi - z_+ - i \varepsilon} \right)  \nonumber\\ & &
~~~~- z_+
\left[\frac{1}{(\xi + z_+ - i \varepsilon)^2} -
             \frac{1}{(\xi - z_+ - i \varepsilon)^2} \right] \Biggr\}
             F(z_+,z_-)~.
\label{eqsym}
\end{eqnarray}
%------------------------------------------------------------------------
Here $F(z_+,z_-)$ 
denotes a distribution amplitude and  $\xi = -q^2/2p.q$~.
$z_+$ and $z_-$ are momentum fractions with
%------------------------------------------------------------------------
\begin{eqnarray}
Dz = dz_+ dz_- \theta(1+z_++z_-) \theta(1+z_+-z_-) \theta(1-z_++z_-)
\theta(1-z_+-z_-) \nonumber\\
\end{eqnarray}
%------------------------------------------------------------------------
The integral over $z_-$ can be performed analytically. Due to the 
structure of the function $F(z_+,z_-)$ the representation
%------------------------------------------------------------------------
\begin{eqnarray}
\int_{-1+|z_+|}^{+1-|z_+|} d z_- F(z_+,z_-) 
= \int_{z_+}^{{\rm sign}(z_+)} \frac{dz}{z} f(z)
\end{eqnarray}
%------------------------------------------------------------------------
holds~\cite{BGR}. Here $f(z)$ denotes the number density of quarks or 
antiquarks. Evidently the structure functions in Eq.~(\ref{eqsym}) 
corresponding to the two tensors are equal but may contain integral
terms w.r.t. a {\it partonic} function. Evaluating the integral
%------------------------------------------------------------------------
\begin{eqnarray}
\lefteqn{
\int_{-1}^{+1} d z_+ 
\frac{z_+}{(\xi \pm z_+ - i \varepsilon)^2}
\int_{z_+}^{{\rm sign}(z_+)} \frac{dz}{z} f(z)}  \nonumber\\ &=&~~~~~~~~~
 \mp
\int_{-1}^{+1} d z_+ 
\frac{1}{\xi \pm z_+ - i \varepsilon} \left[
\int_{z_+}^{{\rm sign}(z_+)} \frac{dz}{z} f(z) - f(z) \right]
\end{eqnarray}
%------------------------------------------------------------------------
the integral terms contributing to the first and second propagator
term in Eq.~(\ref{eqsym}) cancel, while the algebraic term remains.

For the polarized contribution this cancellation does not occur~:
%------------------------------------------------------------------------
\begin{eqnarray}
T_{\mu\nu}^{\rm asym}(q,p) &=& i \varepsilon_{\mu\nu}^{\gamma\beta}
\frac{q_{\gamma} p_{\beta}}{(p.q)^2} q.S \\
& & \times
\int_{-1}^{+1} d z_+ \left[ 
\frac{1}{\xi + z_+ - i \varepsilon} 
+ \frac{1}{\xi - z_+ - i \varepsilon} \right]
\left[\int_{z_+}^{{\rm sign}(z_+)} \frac{dz}{z} f_5(z) - f_5(z)
\right]
\nonumber\\  
& &
- i \varepsilon_{\mu\nu}^{\gamma\beta} \frac{q_{\gamma} S_{\beta}}{p.q}
\int_{-1}^{+1} d z_+ \left[ \frac{1}{\xi + z_+ - i \varepsilon} +
             \frac{1}{\xi - z_+ - i \varepsilon} \right]
\int_{z_+}^{{\rm sign}(z_+)} \frac{dz}{z} f_5(z)~. \nonumber
\end{eqnarray}
%------------------------------------------------------------------------
Rewriting the tensors in terms of those defining the structure functions
$g_1(x,Q^2)$ and $g_2(x,Q^2)$, Eq.~(\ref{had1}),
the Wandzura--Wilczek relation is obtained.
The occurrence of integral contributions or purely algebraic terms
in relations between deeply inelastic structure functions or other hard 
scattering processes has thus to be considered as the regular case, with 
the possibility that these terms may cancel. 
%%%%%%%%%%%%%%%%%%%%%%%%%%%%%%%%%%%%%%%%%%%%%%%%%%%%%%%%%%%%%%%%%%%%%%%%
\section{Twist--3 Relations}
%%%%%%%%%%%%%%%%%%%%%%%%%%%%%%%%%%%%%%%%%%%%%%%%%%%%%%%%%%%%%%%%%%%%%%%%

\vspace{1mm}
\noindent
A unique picture for the twist--3 terms can only be obtained including the
target mass corrections, cf.~\cite{BT}. In this case all polarized
structure functions contain twist--3 contributions. Indeed the 
consideration of mass corrections appears to be necessary also for
consistency reasons since the     structure functions containing
twist--3 terms in the case of longitudinal nucleon polarization are 
weighted by a factor of $M^2/S$ in the scattering cross section.
The explicit relations for the  target mass corrections to
the twist--3 contributions to
$\left. g_i\right|_{i=1}^5$ are given in Ref.~\cite{BT}.

They obey the following relations~:
%----------------------------------------------------------------------
\begin{eqnarray}
 g_1^{\rm III  }(x,Q^2) & = & \frac{4 M^2 x^2}{Q^2}
\left[ g_2^{\rm III  }(x,Q^2)
-2\int_{x}^{1}{\frac{dy}{y} g_2^{\rm III  }(y,Q^2)}\right]~,
\label{t3a}\\
\frac{4 M^2 x^2}{Q^2} g_3^{\rm III  }(x,Q^2) & = &
 g_4^{\rm III  }(x,Q^2)\left(1+\frac{4 M^2 x^2}{Q^2}\right)+
3\int_{x}^{1}{\frac{dy}{y} g_4^{\rm III  }(y,Q^2)}~,
\label{t3b}\\
2 x g_5^{\rm III  }(x,Q^2) &=&
 -\int_{x}^{1}{\frac{dy}{y} g_4^{\rm III  }(y,Q^2)}~,
\label{t3c}
\end{eqnarray}
%---------------------------------------------------------------------------
which hold after the inclusion of the target mass corrections to all
orders in $(M^2/Q^2)^k$.
Whereas the relations (\ref{t3b},\ref{t3c}) are relevant in the presence 
of weak
interactions only, Eq.~(\ref{t3a}) can be tested already for purely 
electromagnetic interactions in the domain of lower values of $Q^2$.

For an experimental determination of the twist--3 contributions to
the structure functions $g_1(x,Q^2)$ and $g_2(x,Q^2)$ one can
proceed as follows. From Eq.~(\ref{t3a}) 
it is evident, that for $Q^2 \gg M^2$
$g_1(x,Q^2)$ receives only twist--2 contributions. $g_1$ can be measured
firstly in this range.              
Its twist--2 contribution at lower values of $Q^2$ can be obtained
solving the twist--2
evolution equations for $g_1(x,Q^2)$. Then one can determine the
twist--2 contribution to $g_2(x,Q^2)$ by the Wandzura--Wilzcek
relation, Eq.~(\ref{eqg2}). 
Assuming that the contributions of twist--4 and
higher are suppressed relatively to the
the twist--3 contributions to $g_{1,2}(x,Q^2)$ the latter can be
extracted form the data by subtraction of the twist--2 pieces.
Relation (\ref{t3a})~\cite{BT} may now be tested by calculating the
twist--3  contribution to $g_1(x,Q^2)$ from the twist--3 contribution to 
$g_2(x,Q^2)$ and comparing with the measurement.

Finally we would like to comment on two other relations.
The Burkhardt--Cottingham sum rule~\cite{BC}
%-----------------------------------------------------------------------
\begin{equation}
\int_0^1 dx g_2(x,Q^2) =0
\end{equation}
%-----------------------------------------------------------------------
is consistent with the results of the local light cone expansion. Both
the corresponding expectation values for twist--2 and twist--3 are
absent in the respective series expansion also in the presence of
target mass corrections~\cite{BT}.

A second relation~\cite{ETL}
%-----------------------------------------------------------------------
\begin{equation}
\int_0^1 dx x\left[g_1(x)+2g_2(x)\right]_{\rm valence}  = 0
\label{eqetl}
\end{equation}
%-----------------------------------------------------------------------
also holds in the presence of target mass  corrections as long as
$Q^2 > M_p^2$. One may cast the respective relations into the following
form, cf.~\cite{BT}:
%-----------------------------------------------------------------------
\begin{equation}
\int_0^1 dx x\left[g_1(x)+2g_2(x)\right]_{\rm valence}
 = \sum_q
\frac{e_q^2}{2} \frac{m_q}
{M_p}
\int_0^1 dx \frac{h_1^q(x)-\overline{h}_1^q(x)}{\displaystyle{
\left(1 - \frac{M^2_px^2}{Q^2}\right)^2}}~.
\label{eqhq}
\end{equation}
%-----------------------------------------------------------------------
Here we allowed for  finite quark masses 
$m_q$ and $h_1^q(x)$ denotes the
transversity distribution of the quark $q$. The integral in 
Eq.~(\ref{eqhq})
is finite under the above condition and one may perform the limit
$m_q \rightarrow 0$ to obtain Eq.~(\ref{eqetl}).
%%%%%%%%%%%%%%%%%%%%%%%%%%%%%%%%%%%%%%%%%%%%%%%%%%%%%%%%%%%%%%%%%%%%%%%%
\section{Summary}
%%%%%%%%%%%%%%%%%%%%%%%%%%%%%%%%%%%%%%%%%%%%%%%%%%%%%%%%%%%%%%%%%%%%%%%%

\vspace{1mm}
\noindent
%-----------------------------------------------------------------------
In the range of low values of $Q^2 \gsim M_p^2$ nucleon mass corrections
to the polarized structure functions in deep inelastic scattering are
essential. After the inclusion of these corrections a symmetric
picture is obtained comparing the respective twist--2 and twist--3
terms, unlike the massless case~\cite{BK1}. In lowest 
order in $\alpha_s$ the
 twist--2 and the twist--3 contributions of the
five polarized structure functions are connected by three relations.
These are the Wandzura--Wilczek relation, a relation by the author and
Kochelev and the Dicus relation for the twist--2 terms, and three 
relations by the author and Tkabladze for the twist--3 terms.
While the Dicus relation receives a finite target mass correction,
the other relations do not, or they do firstly result after the inclusion
of the target mass corrections at all.
The relations being present for purely electromagnetic interactions
can be tested through precision measurements of the structure
functions $g_1(x,Q^2)$ and $g_2(x,Q^2)$ in the near future.

\vspace{2mm}
\noindent
{\bf Acknowledgement.}~I would like to thank the organizers for a very
stimulating symposium and RIKEN for financial support.

%%%%%%%%%%%%%%%%%%%%%%%%%%%%%%%%%%%%%%%%%%%%%%%%%%%%%%%%%%%%%

\end{document}